\documentclass{PoS}
\usepackage{epsfig,float}

\newcommand{\be}{\begin{equation}}
\newcommand{\ee}{\end{equation}}
\newcommand{\ba}{\begin{eqnarray}}
\newcommand{\ea}{\end{eqnarray}}

\title{Interaction of vector mesons with baryons and vectors in the nuclear medium}

\ShortTitle{vector-baryon, vector-nuclei}

\author{\speaker{E. Oset}\thanks{}\\
       Departamento de Fisica Teorica and IFIC, Centro Mixto Universidad de Valencia-CSIC,
Institutos de Investigacion de Paterna, Aptd. 22085, 46071 Valencia, Spain. \\
        E-mail: \email{oset@ific.uv.es}}

\author{A. Ramos\\
       Departament d'Estructura i Constituents de la Materia, Universitat de
Barcelona \\
}

\author{E. J. Garzon\\
      Departamento de Fisica Teorica and IFIC, Centro Mixto Universidad de Valencia-CSIC,
Institutos de Investigacion de Paterna, Aptd. 22085, 46071 Valencia, Spain.\\
}

\author{P. Gonzalez\\
      Departamento de Fisica Teorica and IFIC, Centro Mixto Universidad de Valencia-CSIC,
Institutos de Investigacion de Paterna, Aptd. 22085, 46071 Valencia, Spain.\\
}

\author{J. J. Xie \\
      Departamento de Fisica Teorica and IFIC, Centro Mixto Universidad de Valencia-CSIC,
Institutos de Investigacion de Paterna, Aptd. 22085, 46071 Valencia, Spain.\\
}

\author{A. Martinez Torres \\
      Yukawa Institute for Theoretical Physics, Kyoto University, Kyoto 606-8502, Japan\\
}

\author{L. Tolos \\
      Instituto de Ciencias del Espacio (IEEC/CSIC) Campus Universitat Aut\`onoma de Barcelona, Facultat de Ci\`encies, Torre C5, E-08193 Bellaterra (Barcelona),  Spain\\
}

\author{R. Molina \\
      Departamento de Fisica Teorica and IFIC, Centro Mixto Universidad de Valencia-CSIC,
Institutos de Investigacion de Paterna, Aptd. 22085, 46071 Valencia, Spain.\\
}

\author{C. W. Xiao \\
      Departamento de Fisica Teorica and IFIC, Centro Mixto Universidad de Valencia-CSIC,
Institutos de Investigacion de Paterna, Aptd. 22085, 46071 Valencia, Spain.\\
}

\abstract{ In this talk we present a short review of recent developments concerning the interaction of vector mesons with baryons and with nuclei. 
We begin with the hidden gauge formalism for the interaction of vector mesons, then review results for vector baryon interaction and in particular the resonances which appear as composite states, dynamically generated from the interaction of vector mesons with baryons. New developments concerning the mixing of these states with pseudoscalars and baryons are also reported. We include some discussion on the $5/2^+$ $\Delta$ resonances around 2000 MeV, where we suggest that the $\Delta(2000)5/2^+$ resonance, which comes in the PDG from averaging a set of resonances appearing around 1700 MeV and another one around 2200 MeV, corresponds indeed to two distinct resonances. We also report on a hidden charm baryon state around 4400 MeV coming from the interaction of vector mesons and baryons with charm, and how this state has some repercussion in the  $J/\psi$ suppression in nuclei.  The interaction of $K^*$ in nuclei is also reported and suggestions are made to measure by means of the transparency ratio the huge width in the medium that the theoretical calculations predict. The formalism is extended to $J/\psi$ interaction with nuclei and the transparency ratio for $J/\psi$ photoproduction in nuclei is studied and shown to be a good tool to find possible baryon states which couple to $J/\psi N$.
          }

\FullConference{50th International Winter Meeting on Nuclear Physics - Bormio2012,\\
		23-27 January 2012\\
		Bormio, Italy}

\begin{document}

\section{Introduction}
In this talk we review recent developments on the interaction of vector mesons with baryons and nuclei. For this purpose we use effective field theory using 
a combination of effective Lagrangians to account for hadron interactions,  implementing exactly unitarity in coupled channels. This approach has turned out to be a very efficient tool to 
 face many problems in Hadron Physics. Using this  coupled channel unitary approach, usually referred to as chiral unitary approach, because the Lagrangians used account for chiral symmetry, the 
interaction of the octet of
pseudoscalar mesons with the octet of stable baryons has been studied and 
leads to $J^P=1/2^-$
resonances which fit quite well the spectrum of the known low lying resonances
with these quantum numbers 
\cite{Kaiser:1995cy,angels,ollerulf,carmenjuan,hyodo,ikeda}. 
 New resonances are sometimes predicted, the most notable
being the $\Lambda(1405)$, where all the
chiral approaches find two poles close by 
\cite{Jido:2003cb,Borasoy:2005ie,Oller:2005ig,Oller:2006jw,Borasoy:2006sr,Hyodo:2008xr,Roca:2008kr}, rather than one, for which 
experimental support is presented in \cite{magas,sekihara}.  Another step forward in this
 direction has been the interpretation
of low lying $J^P=1/2^+$ states as molecular systems of two pseudoscalar mesons and one baryon
\cite{alberto,alberto2,kanchan,Jido:2008zz,KanadaEn'yo:2008wm}. 

Much work has been done using pseudoscalar
mesons as building blocks, but more recently, vectors instead of
pseudoscalars are also being considered. In the baryon sector the
interaction of the $\rho \Delta$ interaction has been recently addressed in
\cite{vijande}, where three degenerate $N^*$ states and three degenerate
$\Delta$ states around 1900 MeV, with $J^P=1/2^-, 3/2^-, 5/2^-$, are found. The extrapolation to SU(3) with the interaction of the vectors of the nonet with
the baryons of the decuplet has been done in \cite{sourav}. The
underlying theory for this study is the hidden gauge formalism
\cite{hidden1,hidden2,hidden4}, which deals with the interaction of vector mesons and
pseudoscalars, respecting chiral dynamics, providing the interaction of
pseudoscalars among themselves, with vector mesons, and vector mesons among
themselves. It also offers a perspective on the chiral Lagrangians as limiting
cases at low energies of vector exchange diagrams occurring in the theory.

 In the meson sector, the interaction of $\rho \rho$ within this formalism has
been addressed in \cite{raquel}, where it has been shown to lead  to the
dynamical generation of the $f_2(1270)$ and $f_0(1370)$  meson resonances. The extrapolation to SU(3) of the work 
of \cite{raquel} has been done in \cite{gengvec}, where many resonances are
obtained, some of which can be associated to known meson states, while there are
predictions for new ones.

  In this talk we present the results of the interaction of the nonet
   of vector mesons  with the
   octet of baryons \cite{angelsvec}, which have been obtained
 using the unitary approach in coupled
channels. The scattering amplitudes lead to poles in the
complex plane which can be associated to some well known resonances. Under the
approximation of neglecting the three momentum of the particles versus their
mass, we obtain degenerate states of $J^P=1/2^-,3/2^-$ for the case of the
interaction of vectors with the octet of baryons. This degeneracy 
seems to be followed qualitatively by the experimental spectrum, although in
some cases the spin partners have not been identified. Improvements in the theory follow from the consideration of the decay of these states into a pseudoscalar meson and a baryon, and some results are presented here. 

Moreover we report on composite states of hidden charm emerging from the interaction of vector mesons and baryons with charm.  
We shall also report on some states coming from the three body system 
 $\Delta \rho \pi$ which can shed some light on the status of some $\Delta$ states of $J^P=5/2^+$ in the vicinity of 2000 MeV. Finally, we devote some attention to new developments around vector mesons in a nuclear medium, specifically the $K^*$ and $J/\psi$ in the medium.

\section{Formalism for $VV$ interaction}

We follow the formalism of the hidden gauge interaction for vector mesons of 
\cite{hidden1,hidden2} (see also \cite{hidekoroca} for a practical set of Feynman rules). 
The Lagrangian involving the interaction of 
vector mesons amongst themselves is given by
\begin{equation}
{\cal L}_{III}=-\frac{1}{4}\langle V_{\mu \nu}V^{\mu\nu}\rangle \ ,
\label{lVV}
\end{equation}
where the symbol $\langle \rangle$ stands for the trace in the SU(3) space 
and $V_{\mu\nu}$ is given by 
\begin{equation}
V_{\mu\nu}=\partial_{\mu} V_\nu -\partial_\nu V_\mu -ig[V_\mu,V_\nu]\ ,
\label{Vmunu}
\end{equation}
with  $g$ given by $g=\frac{M_V}{2f}$
where $f=93\,MeV$ is the pion decay constant. The magnitude $V_\mu$ is the SU(3) 
matrix of the vectors of the nonet of the $\rho$
\begin{equation}
V_\mu=\left(
\begin{array}{ccc}
\frac{\rho^0}{\sqrt{2}}+\frac{\omega}{\sqrt{2}}&\rho^+& K^{*+}\\
\rho^-& -\frac{\rho^0}{\sqrt{2}}+\frac{\omega}{\sqrt{2}}&K^{*0}\\
K^{*-}& \bar{K}^{*0}&\phi\\
\end{array}
\right)_\mu \ .
\label{Vmu}
\end{equation}

The interaction of ${\cal L}_{III}$ gives rise to a contact term coming from 
$[V_\mu,V_\nu][V_\mu,V_\nu]$
\begin{equation}
{\cal L}^{(c)}_{III}=\frac{g^2}{2}\langle V_\mu V_\nu V^\mu V^\nu-V_\nu V_\mu
V^\mu V^\nu\rangle\ ,
\label{lcont}
\end{equation}
 and on the other hand it gives rise to a three 
vector vertex from 
\begin{equation}
{\cal L}^{(3V)}_{III}=ig\langle (\partial_\mu V_\nu -\partial_\nu V_\mu) V^\mu V^\nu\rangle
\label{l3V}=ig\langle (V^\mu\partial_\nu V_\mu -\partial_\nu V_\mu
V^\mu) V^\nu\rangle
\label{l3Vsimp}\ ,
\end{equation}

In this latter case one finds an analogy with the coupling of vectors to
 pseudoscalars given in the same theory by 
 
\be
{\cal L}_{VPP}= -ig ~tr\left([
P,\partial_{\mu}P]V^{\mu}\right),
\label{lagrVpp}
\ee
where $P$ is the SU(3) matrix of the pseudoscalar fields. 

In a similar way, we have the Lagrangian for the coupling of vector mesons to
the baryon octet given by
\cite{Klingl:1997kf,Palomar:2002hk}

\be
{\cal L}_{BBV} =
\frac{g}{2}\left(tr(\bar{B}\gamma_{\mu}[V^{\mu},B])+tr(\bar{B}\gamma_{\mu}B)tr(V^{\mu})
\right),
\label{lagr82}
\ee
where $B$ is now the SU(3) matrix of the baryon octet \cite{Eck95,Be95}. Similarly,
one has also a lagrangian for the coupling of the vector mesons to the baryons
of the decuplet, which can be found in \cite{manohar}.

\begin{figure}[tb]
\epsfig{file=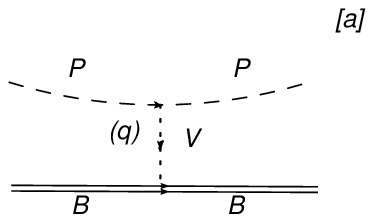, width=7cm} \epsfig{file=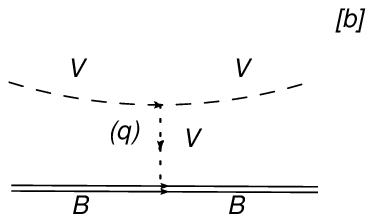, width=7cm}
\caption{Diagrams obtained in the effective chiral Lagrangians for interaction
of pseudoscalar [a] or vector [b] mesons with the octet or decuplet of baryons.}%
\label{f1}%
\end{figure}

With these ingredients we can construct the Feynman diagrams that lead to the $PB
\to PB$ and $VB \to VB$ interaction, by exchanging a vector meson between the
pseudoscalar or the vector meson and the baryon, as depicted in Fig.\ref{f1} .

 As shown in \cite{angelsvec}, in the limit of small three momenta of the vector mesons, which we consider, the vertices of Eq. (\ref{l3Vsimp}) and Eq. (\ref{lagrVpp}) give rise to the same expression.  This makes the work technically easy allowing the use of many previous results.

   A small amendment is in order in the case of vector mesons, which
   is due to the mixing of $\omega_8$ and the singlet of SU(3), $\omega_1$, to give the
   physical states of the $\omega$ and the $\phi$.
    In this case, all one must do is to take the
   matrix elements known for the $PB$ interaction and, wherever $P$ is the
   $\eta_8$, multiply the amplitude by the factor $1/\sqrt 3$ to get the
   corresponding $\omega $ contribution and by $-\sqrt {2/3}$ to get the
   corresponding $\phi$ contribution.  Upon the approximation consistent with
   neglecting the three momentum versus the mass of the particles (in this
   case the baryon), we can just take the $\gamma^0$ component of 
   eq. (\ref{lagr82})  and
   then the transition potential corresponding to the diagram of Fig. 1(b ) is
   given by
   
   \begin{equation}
V_{i j}= - C_{i j} \, \frac{1}{4 f^2} \, (k^0 + k'^0)~ \vec{\epsilon}\vec{\epsilon
} ',
\label{kernel}
\end{equation}
 where $k^0, k'^0$ are the energies of the incoming and outgoing vector mesons. 
   The same occurs in the case of the decuplet.  
    
    The $C_{ij}$ coefficients of Eq. (\ref{kernel}) can be obtained directly from 
    \cite{angels,bennhold,inoue}
    with the simple rules given above for the $\omega$ and the $\phi$, and
    substituting $\pi$ by $\rho$ and $K$ by $K^*$ in the matrix elements. The
    coefficients are obtained both in the physical basis of states or in the
    isospin basis. Here we will show results in isospin
    basis. 

    The next step to construct the scattering matrix is done by solving the
    coupled channels Bethe Salpeter equation in the on shell factorization approach of 
    \cite{angels,ollerulf}
   \begin{equation}
T = [1 - V \, G]^{-1}\, V,
\label{eq:Bethe}
\end{equation} 
with $G$ the loop function of a vector meson and a baryon which we calculate in
dimensional regularization using the formula of \cite{ollerulf} and similar
values for the subtraction constants. The $G$ function is convoluted with the 
spectral function for the vector mesons to take into account their width.

 The iteration of diagrams implicit in the Bethe Salpeter equation in the case
 of the vector mesons propagates the $\vec{\epsilon}\vec{\epsilon }'$ term 
 of the interaction, thus,
the factor $\vec{\epsilon}\vec{\epsilon }'$ appearing in the potential $V$
factorizes also in the $T$ matrix for the external vector mesons. This has as a consequence that the interaction is spin independent and we find degenerate states in $J^P=1/2^-$ and $J^P=3/2^-$.

\section{Results} 

 The resonances obtained are summarized in Table  \ref{tab:octet} .
 As one can see in Table \ref{tab:octet} there are states which one can easily
associate to known resonances. There are ambiguities in other cases. One can also
see that in several cases the degeneracy in spin that the theory predicts is
clearly visible in the experimental data, meaning that there are 
several states with about 50 MeV 
or less mass difference
between them.  In some cases, the theory predicts quantum numbers for 
resonances which have no spin and parity associated. It would be interesting to
pursue the experimental research to test the theoretical predictions.

\begin{table}[ht]
      \renewcommand{\arraystretch}{1.5}
     \setlength{\tabcolsep}{0.2cm}
\begin{tabular}{c|c|cc|ccccc}\hline\hline
$S,\,I$&\multicolumn{3}{c|}{Theory} & \multicolumn{5}{c}{PDG data}\\
\hline
    \vspace*{-0.3cm}
    & pole position    & \multicolumn{2}{c|}{real axis} &  &  & &  &  \\
    & {\small (convolution)}    &\multicolumn{2}{c|}{{\small
    (convolution)}} & \\
    &   & mass & width &name & $J^P$ & status & mass & width \\
    \hline
$0,1/2$ & --- & 1696  & 92  & $N(1650)$ & $1/2^-$ & $\star\star\star\star$ & 1645-1670
& 145-185\\
  &      &       &     & $N(1700)$ & $3/2^-$ & $\star\star\star$ &
	1650-1750 & 50-150\\
%	& & & & & & & \\
       & $1977 + {\rm i} 53$  & 1972  & 64  & $N(2080)$ & $3/2^-$ & $\star\star$ & $\approx 2080$
& 180-450 \\	
   &     &       &     & $N(2090)$ & $1/2^-$ & $\star$ &
 $\approx 2090$ & 100-400 \\
 \hline
$-1,0$ & $1784 + {\rm i} 4$ & 1783  & 9  & $\Lambda(1690)$ & $3/2^-$ & $\star\star\star\star$ &
1685-1695 & 50-70 \\
  &       &       &    & $\Lambda(1800)$ & $1/2^-$ & $\star\star\star$ &
1720-1850 & 200-400 \\
       & $1907 + {\rm i} 70$ & 1900  & 54  & $\Lambda(2000)$ & $?^?$ & $\star$ & $\approx 2000$
& 73-240\\
       & $2158 + {\rm i} 13$ & 2158  & 23  &  &  &  & & \\
       \hline
$-1,1$ & $ --- $ & 1830  & 42  & $\Sigma(1750)$ & $1/2^-$ & $\star\star\star$ &
1730-1800 & 60-160 \\
  & $ --- $   & 1987  & 240  & $\Sigma(1940)$ & $3/2^-$ & $\star\star\star$ & 1900-1950
& 150-300\\
   &     &       &   & $\Sigma(2000)$ & $1/2^-$ & $\star$ &
$\approx 2000$ & 100-450 \\\hline
$-2,1/2$ & $2039 + {\rm i} 67$ & 2039  & 64  & $\Xi(1950)$ & $?^?$ & $\star\star\star$ &
$1950\pm15$ & $60\pm 20$ \\
         & $2083 + {\rm i} 31 $ &  2077     & 29  &  $\Xi(2120)$ & $?^?$ & $\star$ &
$\approx 2120$ & 25  \\
 \hline\hline
    \end{tabular}
\caption{The properties of the 9 dynamically generated resonances stemming from the vector-baryon octet interaction and their possible PDG
counterparts.}
\label{tab:octet}
\end{table}

  The predictions made here for resonances not observed should be a stimulus for
further search of such states. In this
sense it is worth noting the experimental program at Jefferson Lab 
\cite{Price:2004xm} to investigate the $\Xi$ resonances. We are
confident that the predictions  shown here stand on solid grounds and anticipate much
progress in the area of baryon spectroscopy and on the understanding of the
nature of the baryonic resonances. 

\section{Incorporating the pseudoscalar meson-baryon channels}

\begin{center}
\begin{figure}[h!]
%\begin{tabular}{ccc}
\begin{center}
\includegraphics[scale=0.5]{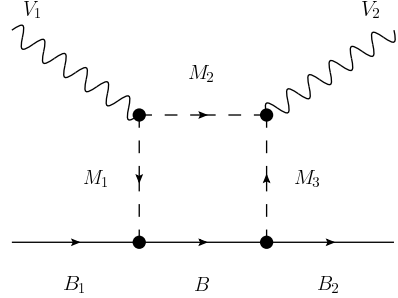} 
\end{center}
%\end{tabular}
\caption{ Diagram for the $VB \rightarrow VB$ interaction incorporating the intermediate pseudoscalar-baryon states.}
\label{box}
\end{figure}
\end{center}

\begin{figure}[ht!]
\begin{center}
\includegraphics[scale=0.5]{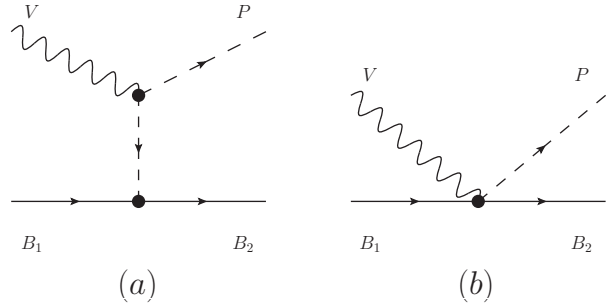}
\end{center}
\caption{Diagrams of the $VB\rightarrow PB$. (a) meson pole term, (b) Kroll Ruderman contact term.}
\label{fig:vbpb}
\end{figure}

Improvements in the states tabulated in Table \ref{tab:octet} have been done by incorporating intermediate states of a pseudoscalar meson and a baryon \cite{garzon}. This is done by including the diagrams of Fig.~(\ref{box}). However, arguments of gauge invariance \cite{Rapp:1997fs,Peters:1997va,Rapp:1999ej,Urban:1999im,Cabrera:2000dx,Cabrera:2002hc,kanchan1,kanchan2} demand that the meson pole term be accompanied by the corresponding Kroll Ruderman contact term, see Fig. \ref{fig:vbpb}.

\begin{figure}[ht!]
\begin{center}
\includegraphics[scale=1.5]{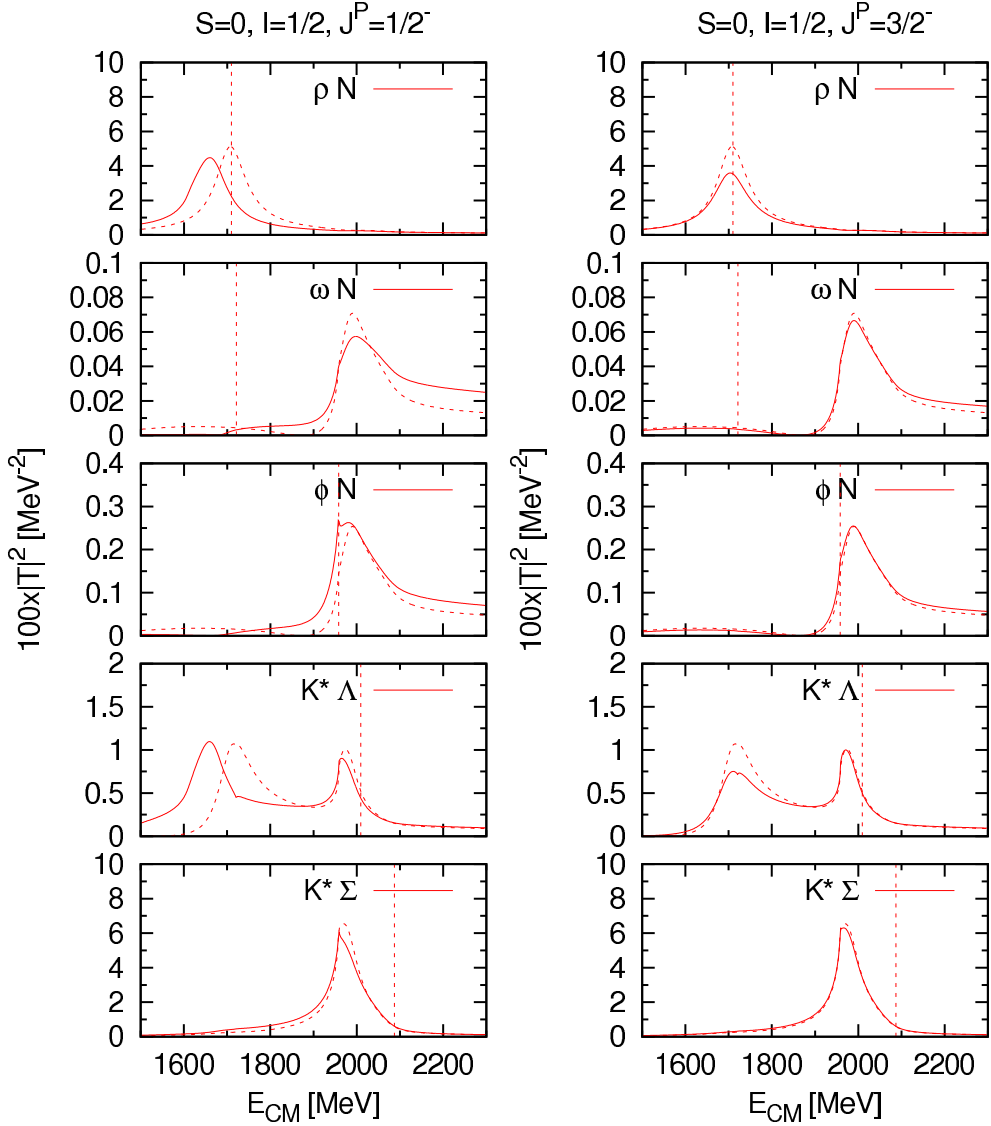} 
\end{center}
\caption{$|T|^2$ for the S=0, I=1/2 states. Dashed lines correspond to tree level only and solid lines are calculated including the box diagram potential. Vertical dashed lines indicate the channel threshold.}
\label{res1}
\end{figure}

In the intermediate B states of Fig. \ref{box} we include baryons of the octet and the decuplet. The results of the calculations are a small shift and a broadening of the resonances obtained with the base of vector-baryon alone.The results are shown in Fig. \ref{res1} and Tables \ref{tab:pdg12} and \ref{tab:pdg32}.  

In Fig.~\ref{res1} we see two peaks for the state of $S=0$ and $I=1/2$, one around 1700 MeV, in channels $\rho N$ and $K^* \Lambda$, and another peak near 1980 MeV, which appears in all the channels except for $\rho N$. 
We can see that the mixing of the PB channels affects differently the two spins, $J^P=1/2^-$ and $3/2^-$, as a consequence of the extra mechanisms contributing to the $J^P=1/2^-$. The effect of the box diagram on the $J^P=3/2^-$ sector is small, however the PB-VB mixing mechanism are more important in the $J^P=1/2^-$ sector. Indeed, the Kroll Ruderman term only allows the $1/2^-$ pseudoscalar baryon intermediate states in the box. The most important feature is a shift of the peak around 1700 MeV, which appears now around 1650 MeV. This breaking of the degeneracy is most welcome since this allows us to associate the $1/2^-$ peak found at 1650 MeV with the $N^*(1650)(1/2^-)$ while the peak for $3/2^-$ at 1700 MeV can be naturally associated to the $N^*(1700)(3/2^-)$. 

%\subsection{I=1/2, S=0}

\begin{table}[ht]
\begin{center}
\begin{tabular}{c|c|cc|ccccc}
\hline\hline
$S,\,I$	&\multicolumn{3}{c|}{Theory} & \multicolumn{5}{c}{PDG data}\\
\hline
    	& pole position	& \multicolumn{2}{c|}{real axis} &  &  & &  &  \\
    	& $M_R+i\Gamma /2$	& mass & width &name & $J^P$ & status & mass & width \\
\hline
$0,1/2$ & $1690+i24^{*}			$	& 1658  & 98  
		& $N(1650)$ & $1/2^-$ & $\star\star\star\star$ 	& 1645-1670	& 145-185\\
  		
       	& $1979+i67			$	& 1973 & 85 
   		& $N(2090)$ & $1/2^-$ & $\star$ 	 & $\approx 2090$ & 100-400 \\
\hline
$-1,0$	& $1776+i39				$	& 1747 & 94  
  		& $\Lambda(1800)$ & $1/2^-$ & $\star\star\star$ 		& 1720-1850 & 200-400 \\
  		
        & $1906+i34^{*}			$ 	& 1890 & 93  
        & $\Lambda(2000)$ & $?^?$ & $\star$ 					& $\approx 2000$ & 73-240\\

        & $2163+i37				$ 	& 2149 & 61  &  &  &  & & \\
\hline
$-1,1$  & $ -			$	& 1829& 84  
		& $\Sigma(1750)$ & $1/2^-$ & $\star\star\star$ & 1730-1800 & 60-160 \\
   		& $ -			 $ 	& 2116 & 200-240  
   		& $\Sigma(2000)$ & $1/2^-$ & $\star$ 			& $\approx 2000$ & 100-450 \\
\hline
$-2,1/2$& $2047+i19^{*}	$	& 2039 & 70  
		& $\Xi(1950)$ & $?^?$ & $\star\star\star$ 	& $1950\pm15$ & $60\pm 20$ \\

        & $ -			$	& 2084 & 53   
        & $\Xi(2120)$ & $?^?$ & $\star$ 			& $\approx 2120$ & 25  \\
\hline\hline
\end{tabular}
\caption{The properties of the nine dynamically generated resonances and their possible PDG
counterparts for $J^P=1/2^-$. The numbers with asterisk in the imaginary part of the pole position are obtained without the convolution for the vector mass distribution of the $\rho$ and $K^*$.}
\label{tab:pdg12}
\end{center}
\end{table}

\begin{table}[ht]
\begin{center}
\begin{tabular}{c|c|cc|ccccc}
\hline\hline
$S,\,I$	&\multicolumn{3}{c|}{Theory} & \multicolumn{5}{c}{PDG data}\\
\hline
    	& pole position	& \multicolumn{2}{c|}{real axis} &  &  & &  &  \\
    	& $M_R+i\Gamma /2$	& mass & width &name & $J^P$ & status & mass & width \\
\hline
$0,1/2$ & $1703+i4^{*}			$	& 1705  & 103  
  		& $N(1700)$ & $3/2^-$ & $\star\star\star$ 		& 1650-1750 & 50-150\\
  		
       	& $1979+i56			$	& 1975 & 72 
       	& $N(2080)$ & $3/2^-$ & $\star\star$ & $\approx 2080$ & 180-450 \\	
\hline
$-1,0$	& $1786+i11				$	& 1785 & 19  
		& $\Lambda(1690)$ & $3/2^-$ & $\star\star\star \star$ 	& 1685-1695 & 50-70 \\
  		
        & $1916+i13^{*}			$ 	& 1914 & 59  
        & $\Lambda(2000)$ & $?^?$ & $\star$ 					& $\approx 2000$ & 73-240\\

        & $2161+i17				$ 	& 2158 & 29  &  &  &  & & \\
\hline
$-1,1$  & $ -			$	& 1839& 58  
  		& $\Sigma(1940)$ & $3/2^-$ & $\star\star\star$ & 1900-1950 & 150-300\\
   		& $ -			 $ 	& 2081 & 270  &  &  &  & & \\  
\hline
$-2,1/2$& $2044+i12^{*}	$	& 2040 & 53  
		& $\Xi(1950)$ & $?^?$ & $\star\star\star$ 	& $1950\pm15$ & $60\pm 20$ \\

        & $2082+i5^{*} 	$	& 2082 & 32   
        & $\Xi(2120)$ & $?^?$ & $\star$ 			& $\approx 2120$ & 25  \\
\hline\hline
\end{tabular}
\caption{The properties of the nine dynamically generated resonances and their possible PDG
counterparts for $J^P=3/2^-$. The numbers with asterisk in the imaginary part of the pole position are obtained without the convolution for the vector mass distribution of the $\rho$ and $K^*$.}
\label{tab:pdg32}
\end{center}
\end{table}

\section{The $\Delta \rho \pi$ system and $\Delta$  $J^P=5/2^+$ states around 2000 MeV}
In Refs.~\cite{vijande,sourav} it was shown that the $\Delta \rho$ interaction
gave rise to $N^*$ and $\Delta$ states with degenerate spin-parity
$1/2^-,3/2^-,5/2^-$. In a recent work~\cite{Xie:2011uw}, one extra
$\pi$ was introduced in the system, and via the Fixed Center
Approximation to the Faddeev Equations, the new system was studied
and new states were found.

\begin{center}%
\begin{table}[ptbh]
\caption{Assignment of $I=3/2$ predicted states to $J^{P}=1/2^{+}%
,3/2^{+},5/2^{+}$ resonances. Estimated PDG masses for these
resonances as well as their extracted values from references
\cite{Man92} and \cite{Vra00} (in brackets) are shown for
comparison. N. C. stands for a non cataloged resonance in the PDG
review}
\begin{tabular}
[c]{c|ccccc} \hline
Predicted & \multicolumn{5}{c}{PDG data}\\
&  &  &  &  & \\
Mass (MeV) & Name & $J^{P}$ & Estimated Mass (MeV) & Extracted Mass (MeV) & Status\\
1770 & $\Delta(1740)$ & $5/2^{+}$ &  & $1752\pm32$ & N.C.\\
&  &  &  & $(1724\pm61)$ & \\
& $\Delta(1600)$ & $3/2^{+}$ & $1550-1700$ & $1706\pm10$ & *** \\
&  &  &  & $(1687\pm44)$ & \\
& $\Delta(1750)$ & $1/2^{+}$ & $\approx1750$ & $1744\pm36$ & * \\
&  &  &  & $(1721\pm61)$ & \\\hline
$1875$ & $\Delta(1905)$ & $5/2^{+}$ & $1865-1915$ & $1881\pm18$ & ****\\
&  &  &  & $(1873\pm77)$ & \\
& $\Delta(1920)$ & $3/2^{+}$ & $1900-1970$ & $2014\pm16$ & ***\\
&  &  &  & $(1889\pm100)$ & \\
& $\Delta(1910)$ & $1/2^{+}$ &
$1870-1920$ & $1882\pm10$ & ****\\
&  &  &  & $(1995\pm12)$ &
\\\hline
\end{tabular}
\label{threebody}
\end{table}
\end{center}

We show in Table \ref{threebody} the two $\Delta^*$ states obtained and
there is also a hint of another $\Delta^*$ state around $2200$ MeV.
Experimentally, only two resonances $\Delta_{5/2^+}(1905)(****)$ and
$\Delta_{5/2^+}(2000)(**)$, are cataloged in the Particle Data Book
Review~\cite{pdg2010}. However, a careful look at
$\Delta_{5/2^+}(2000)(**)$, shows that its nominal mass is in fact
estimated from the mass $(1724\pm61)$, $(1752\pm32)$ and
$(2200\pm125)$ respectively, extracted from three independent
analyses of different character~\cite{Man92,Vra00,Cut80}. Moreover a
recent new data analysis~\cite{Suz10} has reported a
$\Delta_{5/2^+}$ with a pole position at $1738$ MeV.

Our results give quantitative theoretical support to the existence
of two distinct $5/2^+$ resonances, $\Delta_{5/2^+}(\sim 1740)$ and
$\Delta_{5/2^+}(\sim 2200)$, apart from the one around 1905 MeV . We propose that these two resonances
should be cataloged instead of $\Delta_{5/2^+}(2000)$. This proposal
gets further support from the possible assignment of the other
calculated baryon states in the $I=1/2,3/2$ and $J^P=1/2^+,3/2^+$
sectors to known baryonic resonances. In particular the poorly
established $\Delta_{1/2^+}(1750)(*)$ may be naturally interpreted
as a $\pi N_{1/2^-}(1650)$ bound state.

\section{Hidden charm baryons from vector-baryon interaction}
Following the idea of \cite{angelsvec} in \cite{wu} is was found that several baryon states emerged as hidden charm composite states of mesons and baryons with charm. In particular along the lines discussed here, we find a hidden charm baryon that couples to $J/\psi N$ and other channels, as shown in Tables \ref{jpsicoupling},\ref{jpsiwidth}.  This will play a role later on when we discuss the $J/\psi$ suppression in nuclei. To do the calculations in the charm sector an extension to SU(4) of the hidden gauge Lagrangians is made, but the symmetry is explicitly broken when considering the exchange of heavy vector mesons, where the appropriate reduction in the Feynman diagrams is taken into account. 

                                                                                   \begin{table}[ht]
      \renewcommand{\arraystretch}{1.1}
     \setlength{\tabcolsep}{0.4cm}
\begin{center}
\begin{tabular}{cccccc}\hline
$(I, S)$&  $z_R$                   & \multicolumn{4}{c}{$g_a$}\\
\hline
$(1/2, 0) $   &             & $\bar{D}^{*} \Sigma_{c}$ & $\bar{D}^{*} \Lambda^{+}_{c}$&  $J/\psi N$ \\
          & $4415-9.5i$   & $2.83-0.19i     $          &$-0.07+0.05i   $            &  $-0.85+0.02i$ \\
          &             &$ 2.83  $                 &$0.08  $                  &  $0.85$ \\
\hline
$(0, -1) $ &                & $\bar{D}^{*}_{s} \Lambda^{+}_{c}$   & $\bar{D}^{*} \Xi_{c}$ & $\bar{D}^{*} \Xi'_{c}$ & $J/\psi \Lambda$\\
       &   $4368-2.8i  $  & $1.27-0.04i     $                     &$ 3.16-0.02i $          & $-0.10+0.13i  $          & $0.47+0.04i   $     \\
       &                & $1.27 $                             & $3.16 $               & $0.16 $                & $0.47  $          \\
       &   $4547-6.4i $   & $0.01+0.004i$                         & $0.05-0.02i$            & $2.61-0.13i $            & $-0.61-0.06i   $                \\
       &                & $0.01   $                           & $0.05 $               & $2.61$                 & $0.61 $                  \\
\hline\end{tabular} \caption{Pole position ($z_R$) and coupling
constants ($g_a$) to various channels for the states from
$PB\rightarrow PB$ including the $J/\psi N$ and $J/\psi\Lambda$
channels. }
\label{jpsicoupling}
\end{center}
%\end{table}
%\begin{table}[ht]
       \renewcommand{\arraystretch}{1.1}
     \setlength{\tabcolsep}{0.4cm}
\begin{center}
\begin{tabular}{ccccc}\hline
$(I, S)$      &  $z_R$      & \multicolumn{2}{c}{Real axis} & $\Gamma_i$ \\
          &    & $M$ & $\Gamma$                  & \\
\hline
$(1/2, 0)$    &            &      &             &  $J/\psi N$\\
          & $4415-9.5i$  & $4412$ & $47.3$        &  $19.2$       \\
\hline
$(0, -1)$     &            &      &             &  $J/\psi\Lambda$\\
          & $4368-2.8i$  & $4368 $& $28.0 $       &  $5.4$        \\
          & $4547-6.4i $ & $4544 $& $36.6 $       &  $13.8$        \\
\hline\end{tabular} \caption{Pole position ($z_R$), mass ($M$),
total width ($\Gamma$, including the contribution from the light
meson and baryon channel) and the decay widths for the $J/\psi N$
and $J/\psi\Lambda$ channels ($\Gamma_i$). The unit are in MeV}
\label{jpsiwidth}
\end{center}
\end{table}  

\section{The properties of $K^*$ in nuclei}
Much work about the vector mesons in nuclei has been done in the last decade
\cite{rapp,hayano,mosel,na60,wood,nucl-th/0610067,nanova}, mostly for the $\rho,\phi,\omega$, which can be studied looking for dileptons. Maybe this technical detail was what prevented any attention being directed to the renormalization of the  $K^*$ in nuclei. However, recently this work has been addressed in \cite{lauraraquel} with very interesting results.

%\vspace{0.3cm}

The $K^{*-}$ width in vacuum is determined by the imaginary part of the free $\bar K^*$ self-energy at rest,  ${\rm Im} \Pi^0_{\bar K^*}$, due to the decay of the $\bar{K}^*$ meson into $\bar{K}\pi$ pairs:  $\Gamma_{K^{*-}}=-\mathrm{Im}\Pi_{\bar{K}^*}^{0}/m_{\bar K^*}=42$ MeV \cite{lauraraquel}. Note that this value is quite close to the experimental value $\Gamma^{\rm exp}_{K^{*-}}=50.8\pm 0.9$ MeV.

\begin{figure}[t]
\begin{center}
\includegraphics[width=0.45\textwidth,height=5cm]{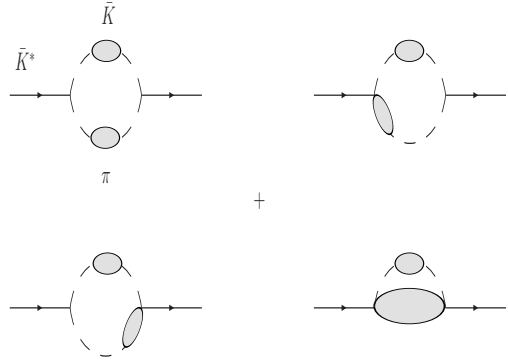}
\hfill
\caption{Self-energy diagrams from the decay of the $\bar{K}^*$ meson in the medium.}
\label{fig:1}
\end{center}
\end{figure}

\begin{figure}[ht]
\begin{center}
\includegraphics[width=0.8\textwidth]{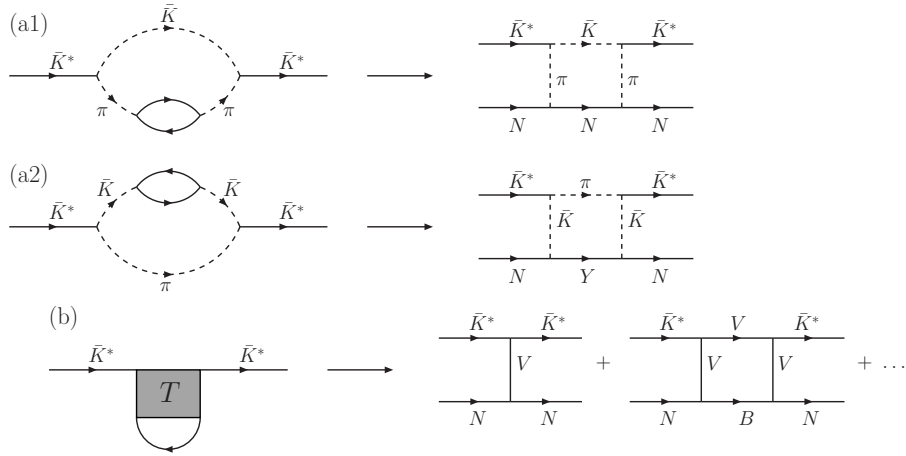}
\caption{Contributions to the $\bar K^*$ self-energy, depicting their different
inelastic sources.}
\label{fig:3}
\end{center}
\end{figure}

The $\bar{K}^*$ self-energy  in matter, on one hand, results from its decay into ${\bar K}\pi$, $\Pi_{\bar{K}^*}^{\rho,{\rm (a)}}$, including both the self-energy of the antikaon \cite{Tolos:2006ny} and the pion \cite{Oset:1989ey,Ramos:1994xy} (see first diagram of Fig.~\ref{fig:1} and some specific contributions in diagrams $(a1)$ and $(a2)$ of Fig.~\ref{fig:3}). Moreover, vertex corrections required by gauge invariance are also incorporated, which are associated to the last three diagrams in  Fig. \ref{fig:1}.

%\subsection{$\bar K^*$ self-energy from s-wave $\bar K^* N$ interaction in
%matter}
%\vspace{0.3cm}

The second contribution to the $\bar K^*$ self-energy comes from its interaction
with the nucleons in the Fermi sea, as displayed in diagram (b) of 
Fig.~\ref{fig:3}. This accounts for the direct quasi-elastic process $\bar K^* N \to \bar K^* N$ as well as other absorption channels $\bar K^* N\to \rho Y, \omega Y, \phi Y, \dots$ with $Y=\Lambda,\Sigma$. 
This contribution is determined by
integrating the medium-modified $\bar K^* N$ amplitudes, $T^{\rho,I}_{\bar
K^*N}$, over the  Fermi sea of nucleons,
\begin{eqnarray}
\Pi_{\bar{K}^*}^{\rho,{\rm (b)}}(q^0,\vec{q}\,)&=&\int \frac{d^3p}{(2\pi)^3} \, n(\vec{p}\,)\,
\left [~{T^\rho}^{(I=0)}_{\bar K^*N}(P^0,\vec{P})+3 {T^\rho}^{(I=1)}_{\bar K^*N}(P^0,\vec{P})\right ] \ ,
 \label{eq:pid}
\end{eqnarray}
where $P^0=q^0+E_N(\vec{p}\,)$ and $\vec{P}=\vec{q}+\vec{p}$ are the
total energy and momentum of the $\bar K^*N$ pair in the nuclear
matter rest frame, and the values $(q^0,\vec{q}\,)$ stand for the
energy and momentum of the $\bar K^*$ meson also in this frame. The
self-energy $\Pi_{\bar{K}^*}^{\rho,{\rm (b)}}$ has to be determined self-consistently
since it is obtained from the in-medium amplitude
${T}^\rho_{\bar K^*N}$ which contains the $\bar K^*N$ loop function
${G}^\rho_{\bar K^*N}$, and this last quantity itself is a function of the complete self-energy
$\Pi_{\bar K^*}^{\rho}=\Pi_{\bar{K}^*}^{\rho,{\rm (a)}}
+\Pi_{\bar{K}^*}^{\rho,{\rm (b)}}$.

%\subsection{Comments on the $\bar K^*$ self-energy in matter}

We note that the two contributions to the $\bar K^*$ self-energy, coming from
the decay of
$\bar K \pi$ pairs in the medium [Figs.~\ref{fig:3}(a1) and \ref{fig:3}(a2)] or
from the  $\bar K^* N$ interaction [Fig.~\ref{fig:3}(b)] provide different
sources
of inelastic $\bar K^* N$ scattering, which add incoherently in the $\bar K^*$
width.  
As seen in the upper two rows of Fig.~\ref{fig:3}, the  $\bar K^* N$
amplitudes mediated by intermediate $\bar K N$ or $\pi
Y$ states are not unitarized. Ideally, one would like to treat the vector
meson-baryon ($VB$) and pseudoscalar meson-baryon ($PB$) states on the same
footing. However, at the energies of interest, transitions of the type $\bar K^*
N \to \bar K N$ mediated by pion exchange may place this pion on its mass shell,
forcing one to keep track of the proper analytical cuts contributing to the
imaginary part of the amplitude and making the iterative process more
complicated. A technical solution can be found by calculating the box diagrams
of Figs.~\ref{fig:3}(a1) and \ref{fig:3}(a2), taking all the cuts into account
properly, and adding the resulting $\bar K^* N \to \bar K^* N$ terms to the $VB
\to V^\prime B^\prime$ potential coming from vector-meson exchange, in a
similar way as done for the study of the vector-vector interaction in
Refs.~\cite{raquel,gengvec}. As we saw in the former sections, the generated resonances barely change their position for spin 3/2 and only by a moderate amount in some cases for spin 1/2. Their widths are somewhat enhanced due to the opening of the newly
allowed $PB$ decay channels \cite{garzon}.

\begin{figure}[ht]
\begin{center}
\includegraphics[width=0.45\textwidth]{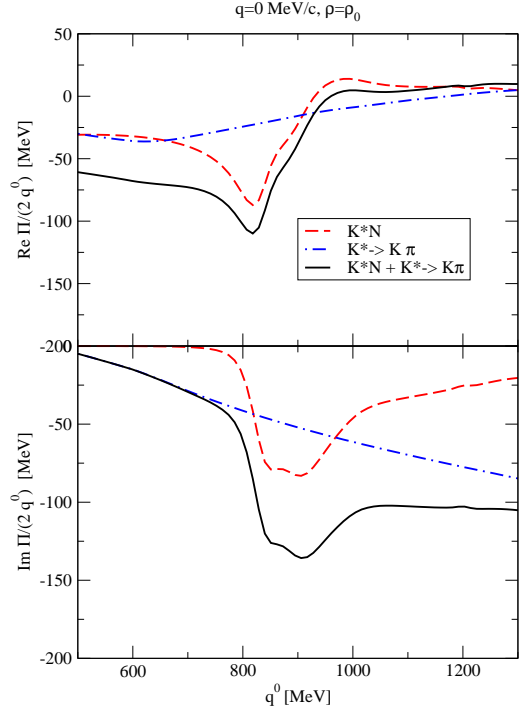}
\hfill
\caption{  $\bar K^*$ self-energy for
 $\vec{q}=0 \, {\rm MeV/c}$ and $\rho_0$. }
\label{fig:auto-spec}
\end{center}
\end{figure}

The full $\bar K^*$ self-energy as a function of the $\bar K^*$ energy for zero
momentum at normal nuclear matter density is shown  in 
Fig.~\ref{fig:auto-spec}. We explicitly indicate the contribution to the
self-energy coming from the self-consistent calculation of the $\bar K^* N$
effective interaction (dashed lines) and the self-energy from the $\bar K^*
\rightarrow \bar K \pi$ decay mechanism (dot-dashed lines), as well as the
combined result from both sources (solid lines).

Around $q^0= 800-900$ MeV we observe an enhancement of the width as well as some structures in the real part of the $\bar K^*$ self-energy. The origin of these structures can be traced back to  the coupling of the $\bar K^*$ to the in-medium $\Lambda(1783) N^{-1}$ and  $\Sigma(1830) N^{-1}$ excitations, which dominate the $\bar K^*$ self-energy in this energy region. However, at lower energies where the $\bar K^* N\to V B$ channels 
are closed, or at large energies beyond the resonance-hole excitations,
the width of the $\bar K^*$ is governed by the $\bar K \pi$ decay mechanism in dense matter.

As we can see, the $\bar K^*$ feels a moderately attractive optical potential and acquires a width of $260$ MeV, which is about five times its width in vacuum. In the next section we devise a method to measure this large width experimentally.

\section{Transparency ratio for  $\gamma A \to K^+ K^{*-} A'$}
%\vspace{0.5cm}

The width of the $\bar K^*$ meson in nuclear matter can be analyzed experimentally by means of the nuclear transparency ratio. The aim is to compare the cross sections of the photoproduction reaction $\gamma A \to K^+ K^{*-} A'$ in different nuclei, and tracing the differences to the in medium $K^{*-}$ width.

\begin{figure}[ht]
\begin{center}
\includegraphics[width=0.6\textwidth]{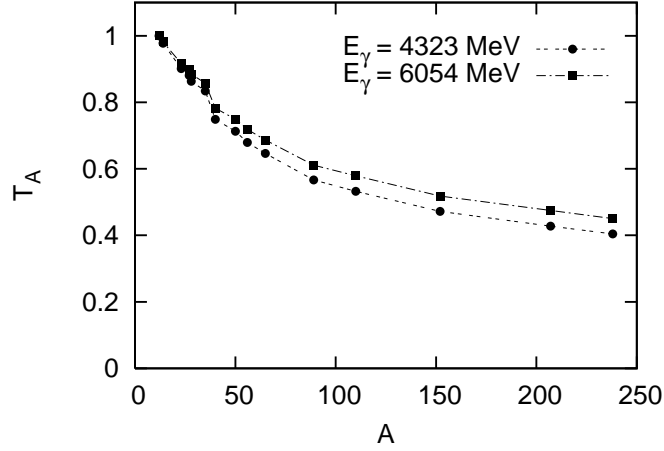}
\caption{Nuclear transparency ratio  for  $\gamma A \to K^+ K^{*-} A'$ for
different nuclei}
\label{fig:ratio}
\end{center}
\end{figure}

The normalized nuclear transparency ratio is defined as
\begin{equation}
T_{A} = \frac{\tilde{T}_{A}}{\tilde{T}_{^{12}C}} \hspace{1cm} ,{\rm with} \ \tilde{T}_{A} = \frac{\sigma_{\gamma A \to K^+ ~K^{*-}~ A'}}{A \,\sigma_{\gamma N \to K^+ ~K^{*-}~N}} \ .
\end{equation}
The quantity $\tilde{T}_A$ is the ratio of the nuclear $K^{*-}$-photoproduction cross section
divided by $A$ times the same quantity on a free nucleon. This describes the loss of flux of $K^{*-}$ mesons in the nucleus and is related to the absorptive part of the $K^{*-}$-nucleus optical potential and, therefore, to the $K^{*-}$ width in the nuclear medium.  In order to remove other nuclear effects not related to the absorption of the $K^{*-}$, we evaluate this ratio with respect to $^{12}$C, $T_A$.

In  Fig.~\ref{fig:ratio} we show the transparency ratio for different nuclei and for two energies in the center of mass reference system, $\sqrt{s}=3$ GeV and $3.5$ GeV, or, equivalently, two energies of the photon in the lab frame of $4.3$ GeV and $6$ GeV respectively. There is a very strong attenuation of the $\bar{K}^*$ survival probability coming from the decay or absorption channels $\bar{K}^*\to \bar{K}\pi$ and $\bar{K}^*N\to \bar K^* N, \rho Y, \omega Y, \phi Y, \dots$, with increasing nuclear-mass number $A$. The reason is the larger path that the $\bar{K}^*$ has to follow before it leaves the nucleus, having then more chances of decaying or being absorbed.

\section{$J/\psi$ suppression}
$J/\psi$ suppression in nuclei has been a hot topic \cite{Vogt:1999cu}, among others for its possible interpretation as a signature of the formation of quark gluon plasma in heavy ion reactions \cite{Matsui:1986dk}, but many other interpretations have been offered  \cite{Vogt:2001ky,Kopeliovich:1991pu,Sibirtsev:2000aw}. In a recent paper \cite{raquelxiao} a study has been done of different $J/\psi N$ reactions which lead to $J/\psi$ absorption in nuclei. The different reactions considered are 
the transition of  $J/\psi N$ to $VN$ with $V$ being a light vector, $\rho, \omega,\phi$, together with the inelastic channels, 
$J/\psi N \to \bar D \Lambda_c$ and $J/\psi N \to \bar D \Sigma_c$. 
Analogously, we consider the mechanisms where the exchanged $D$ collides with a nucleon and gives $\pi \Lambda_c$ or $\pi \Sigma_c$. The cross section has a peak around $\sqrt s=4415$ MeV, where the $J/\psi N$ couples to the resonance described in Tables \ref{jpsicoupling} and \ref{jpsiwidth}. With the inelastic cross section obtained we study the transparency ratio for electron induced $J/\psi$ production in nuclei at about 10 GeV and find that 30 - 35 \% of the $J/\psi$ produced in heavy nuclei are absorbed inside the nucleus. This ratio is in line with depletions of $J/\psi$ in matter observed in other reactions. This offers a novelty in the interpretation of the $J/\psi$ suppression in terms of hadronic reactions, which has also been advocated before \cite{Sibirtsev:2000aw}. Apart from novelties in the details of the calculations and the reaction channels considered, we find that the presence of the resonance that couples to $J/\psi N$ produces a peak in the inelastic $J/\psi N$ cross section and a dip in the transparency ratio. A measure of this magnitude could lead to an indirect observation of the existence of resonances that couple to $J/\psi N$. 

\begin{figure}[ht]
\begin{center}
\includegraphics[width=0.6\textwidth]{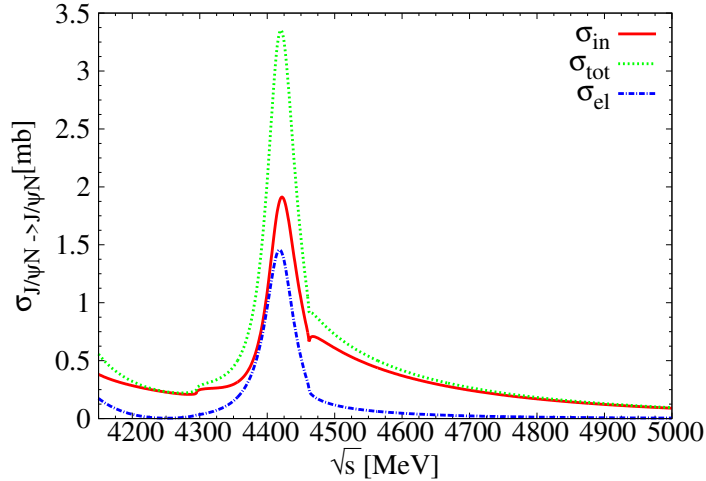}
\caption{The total, elastic and inelastic cross sections in $J/\psi N$ to $J/\psi N$ through intermediate vector-baryon states.}\label{crosec}
\end{center}
\end{figure}

In Fig. \ref{crosec} we can see the total, elastic and inelastic cross sections in $J/\psi N$ to $J/\psi N$ through intermediate vector-baryon states. We can clearly see the peak around 4415 MeV produced by the hidden charm resonance dynamically generated form the interaction of $J/\psi N$ with its coupled $VB$ channels. 

\begin{figure}[ht]
\begin{center}
\includegraphics[width=0.5\textwidth]{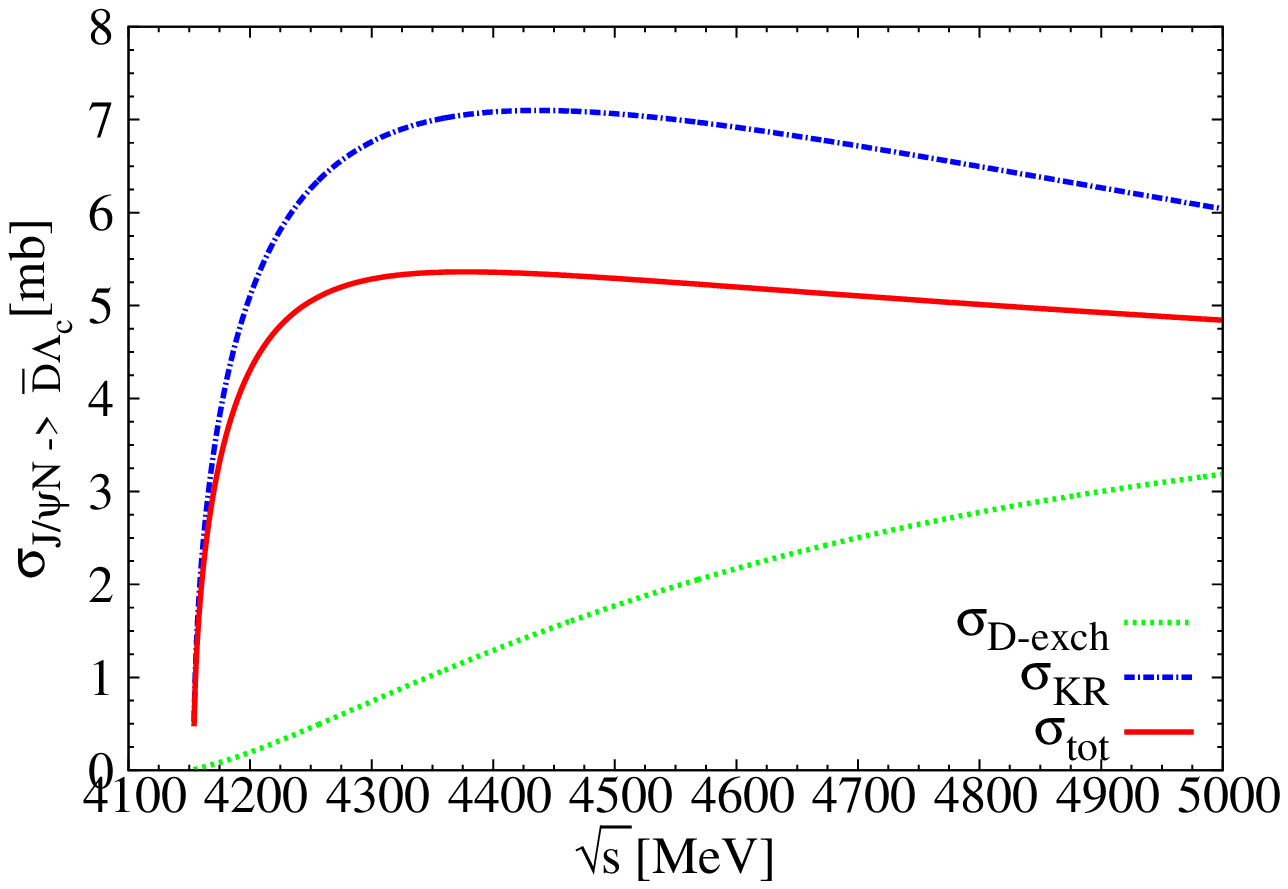}
\includegraphics[width=0.5\textwidth]{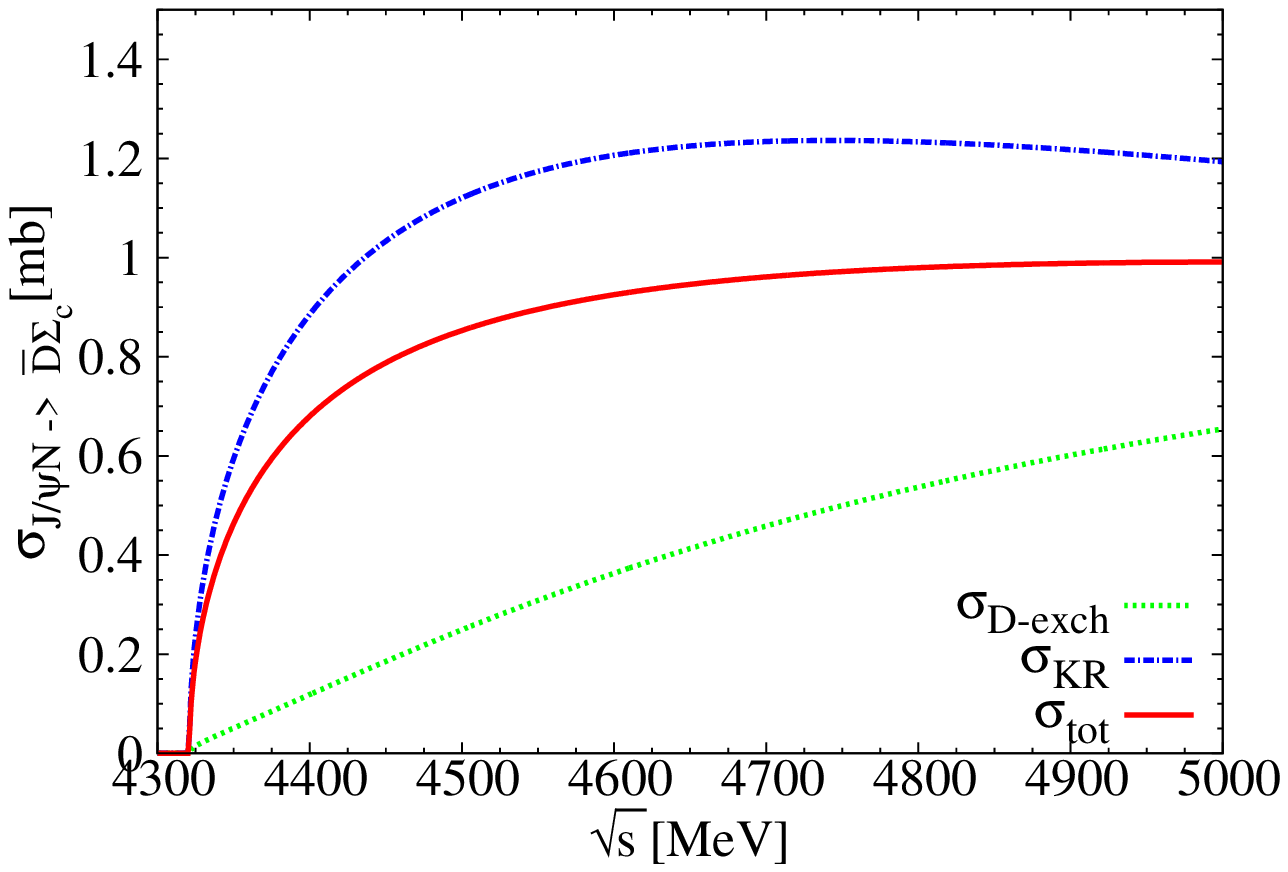}
\caption{The cross section for $J/\psi N \to \bar{D} \Lambda_c$ (up) and $J/\psi N\to \bar{D} \Sigma_c$ (down).}\label{figcro}
\end{center}
\end{figure}

In Fig. \ref{figcro} we can see the cross section for $J/\psi N \to \bar{D} \Lambda_c$  and $J/\psi N\to \bar{D} \Sigma_c$. We can see that numerically the first cross section is sizeable, bigger than the one from the $VB$ channels.

The cross sections for $J/\psi N$ to $\bar D \pi \Lambda_c$ or $\bar D \pi \Sigma_c$ are small in size in the region of interest and are not plotted here. 

\begin{figure}[ht]
\begin{center}
\includegraphics[width=0.6\textwidth]{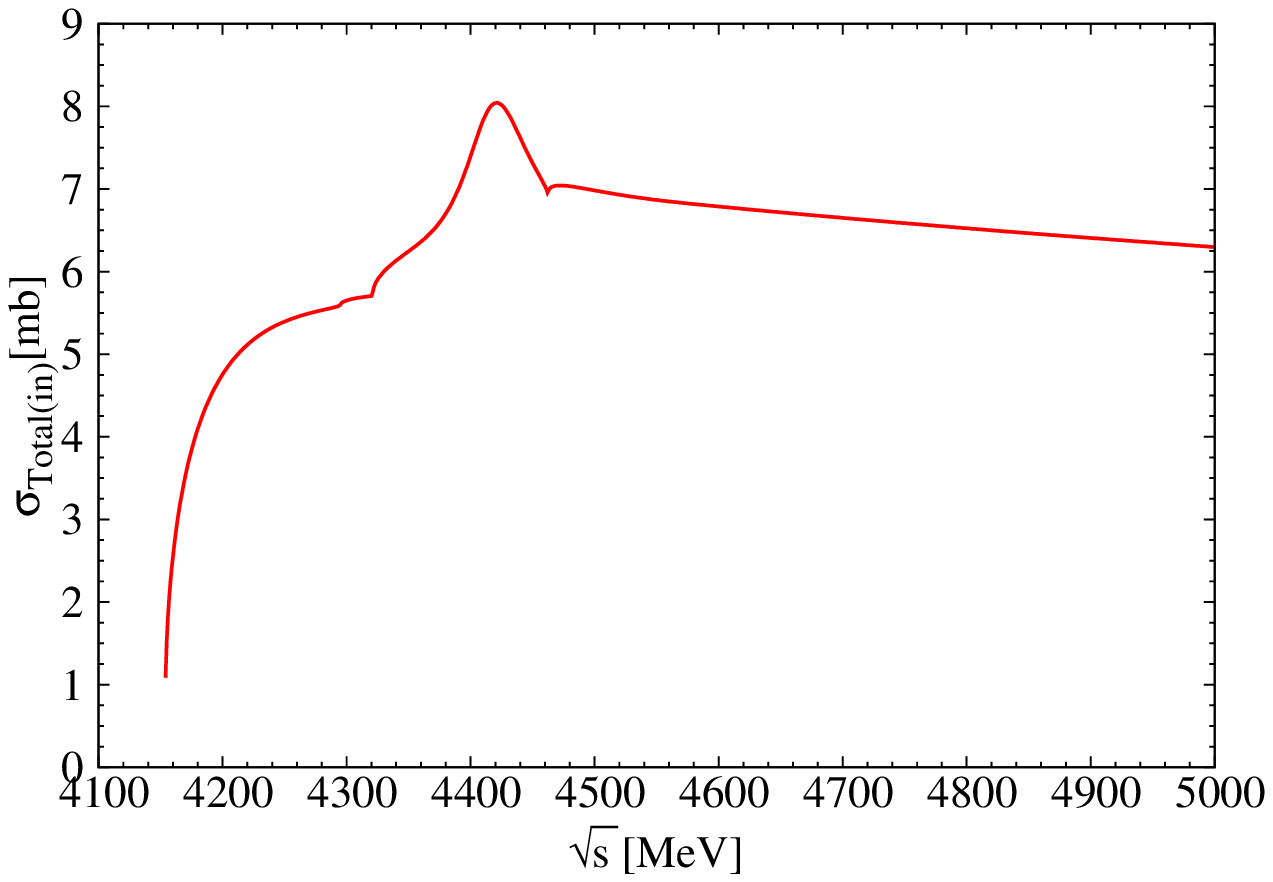}
\caption{The total inelastic cross section of $J/\psi N$.}\label{sigin}
\end{center}
\end{figure}

In Fig. \ref{sigin} we plot  the total $J/\psi N$ inelastic cross section, as the sum of all inelastic cross sections from the different sources discussed before.

\begin{figure}[ht]
\begin{center}
\includegraphics[width=0.6\textwidth]{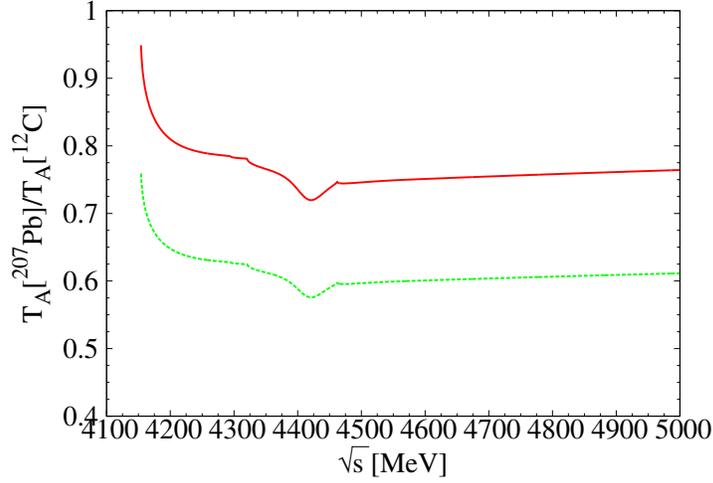}
\caption{The transparency ratio of $J/\psi$ photoproduction  as a function of the energy in the CM of $J/\psi$ with nucleons of the nucleus. Solid line: represents  the effects due to $J/\psi$ absorption. Dashed line: includes photon shadowing \cite{Bianchi:1995vb}.}\label{transg}
\end{center}
\end{figure}

Finally, in Fig. \ref{transg} we show the transparency ratio of $^{208}Pb$ versus the one of $^{12}C$ as a function of the energy. We find sizeable reductions in the rate of $J/\psi$ production in electron induced reactions, and particularly a dip in the ratio that could be searched experimentally. 
It should be noted that the calculation of the transparency ratio done so far does not consider the shadowing of the photons and assumes they can reach every point without being absorbed. However, for $\gamma$ energies of around 10 GeV, as suggested here, the photon shadowing cannot be ignored. Talking it into account is easy since one can multiply the ratio $T_A'$ by the ratio of $N_{eff}$ for the nucleus of mass $A$ and $^{12}C$. This ratio for $^{208}Pb$ to $^{12}C$ at $E_\gamma =$10 GeV is of the order $0.8$ but with uncertainties \cite{Bianchi:1995vb}. We should then multiply $T_A'(^{208}Pb)$ in Fig. \ref{transg} by this extra factor for a proper comparison with experiment.
Although a good resolution should be implemented, the prospect of finding new states through this indirect method should serve as a motivation to perform such experiments.

\section{Conclusions}

We have made a survey of recent developments along the interaction of vector mesons with baryons and the properties of some vector mesons in a nuclear medium.  We showed that the interaction is strong enough to produce resonant states which can qualify as quasibound states of a vector meson and a baryon in coupled channels. This adds to the wealth of composite states already established from the interaction of pseudoscalar mesons with baryons. At the same time we offered the results of new pictures that mix the pseudoscalar-baryon states with the vector-baryon states and break the spin degeneracy that the original model had.  The method of vector-baryon interaction extended to the charm sector also produced some hidden charm states which couple to the $J/\psi N$  channel and had some repercussion in the $J/\psi$ suppression in nuclei. We also showed results for the spectacular renormalization of the $K^*$ in nuclei, where the width becomes as large as 250 MeV at normal nuclear matter density and we made suggestions of experiments that could test this large change.

\section*{Acknowledgments}  
This work is partly supported by DGICYT contract number
FIS2011-28853-C02-01, the Generalitat Valenciana in the program Prometeo, 2009/090. L.T. acknowledges support from Ramon y Cajal Research Programme, and from FP7-PEOPLE-2011-CIG under contract PCIG09-GA-2011-291679. We acknowledge the support of the European Community-Research Infrastructure
Integrating Activity
Study of Strongly Interacting Matter (acronym HadronPhysics3, Grant Agreement
n. 283286)
under the Seventh Framework Programme of EU.

\end{document}